# First performance evaluation of a Multi-layer Thick Gaseous Electron Multiplier with in-built electrode meshes - MM-THGEM

**Rui de Olivera[a] and Marco Cortesi[b]**

[a] CERN,
  Geneva, Switzerland

[b] National Superconducting Cyclotron Laboratory,
  East Lansing (MI), USA
  E-mail: cortesi@nscl.msu.edu

ABSTRACT: We describe a new micro-pattern gas detector structure comprising a multi-layer hole-type multiplier (M-THGEM) combined with two in-built electrode meshes: the Multi-Mesh THGEM-type multiplier (MM-THGEM). Suitable potential differences applied between the various electrodes provide an efficient collection of ionization electrons within the MM-THGEM holes and a large charge avalanche multiplication between the meshes. Different from conventional hole-type multipliers (e.g. Gas Electron Multipliers - GEMs, Thick Gas Electron Multipliers – THGEMs, etc.), which are characterized by a variable (dipole-like) field strength inside the avalanche gap, electrons in MM-THGEMs are largely multiplied by a strong uniform field established between the two meshes, like in the parallel-plate avalanche geometry. The presence of the two meshes within the holes allows for the trapping of a large fraction of the positive ions that stream back to the drift region. A gas gain above $10^5$ has been achieved for single photo-electron detection with a single MM-THGEM in Ar/(10%)$CH_4$ and He/(10%)$CO_2$, at standard conditions for temperature and pressure. When the MM-THGEM is coupled to a conventional THGEM and used as first cascade element, the maximum achievable gains reach values above $10^6$ in He/(10%)$CO_2$, while the IBF approaches of 1.5% in the case of optimum detector-bias configuration. This IBF value is several times lower compared to the one obtained by a double GEM/THGEM detector (5-10%), and equivalent to the performance attained by a Micromegas detector.

# Contents

## 1. Introduction

Since the beginning of the current century, one of the main trends in Micro-Pattern Gaseous Detectors (MPGDs) focuses on the development of brand-new gas avalanche structures [1]. The ultimate goal is to find alternative solutions able to further overcome the limitations of the current MPGD architectures and to fulfil the most stringent constraints imposed by future facilities. This includes operation in harsh experimental conditions as well as increasingly higher operational performance.

As a result, the search for an efficient suppression of the ion-backflow (IBF) is an extremely active field of research, driven by development of Gaseous Photo-Multipliers – GPMs [1], [2], as well as by the spread of continuous-mode Time Projection Chambers (TPC) applications in high energy physics and in other areas [3], [4]. A low level of IBF allows one to reduce secondary effects (i.e. secondary electron emissions) and the photocathode aging induced by ion impacts [5] in GPMs. In addition, it prevents space-charge field distortion induced by ion migration into the drift volume in TPCs, which causes significant loss of the event topology resolution [6]. Finally, a lower amount of ions impinging on the cathode surface may also reduce the probability of sporadic discharges due to cathode-excitation effect [7], particularly at high rate.

For a single avalanche multiplier, i.e. a single Gas Electron multiplier (GEM) [8] or a MICRO-MEsh-GAseous Structure (Micromegas) [9], the IBF is proportional to the ratio between the drift and the avalanche fields, while for multiple hole-type architectures low IBF is achieved by a suitable gas gain sharing between the various cascade elements, as well as an accurate optimization of the transfer fields between them. Micromegas detectors offer a natural suppression of the IBF by efficiently trapping the slow ions on the mesh, leading to IBF values below 1% in normal operational conditions [10]. Typical IBFs of the order of 5%-10% are achieved in two-cascade GEM/thick-GEM [11] detectors, in standard gas mixtures (i.e. Ar-based mixtures) and at gain of $10^3$-$10^4$ [11], [12]. A further lower IBF value (down below 5%) can be achieved in a multi-element (two/three) THGEM/GEM detector by staggering the hole alignment between the various multipliers [13].

The Micro-Strip Hole Plate (MSHP) [14], the COBRA [15] and the thick-COBRA [16] are examples of hybrid MPGD designs formed in a single structure that combine micro-hole electron multiplier geometry with patterned electrode surfaces for ion defocusing. A cascade of MSHPs (or thick-COBRA) was reported to reach ion blocking at the level of $10^{-2}$% in DC mode [15].

Even lower IBF values of around $10^{-4}$% at a gain of $10^5$ were attained with multi-elements detector comprising two GEMs sandwiched between two COBRA multipliers [15]. However, the major drawback of MHSPs (as well as of the thick-COBRA structures) is the loss of effective area on its top surface, where the reflective photocathode is deposited. In addition, the distortion of the dipole field geometry on the top surface of the MHSPs due to the presence of the reverse-biased strip electrodes contributes to a significant loss of electron collection efficiency [13]. A similar effectual IBF reduction of close to few $10^{-4}$% for gains of a several thousand has also been attained with double-mesh Micromegas devices, thought at the expense of a low electron transfer efficiency (computed as only 40%) [17]. Double-mesh Micromegas are presently limited in size to a few tens of cm$^2$.

This work presents and discusses a novel micro-pattern gaseous detector structure (the Multi-Mesh THGEM-type multiplier – MM-THGEM) that merges the double-mesh Micromegas design [17] and the robust multi-layer THGEM [18]. The former serves for avalanche amplification and IBF suppression, while the latter provides a rigid support for large area coverage. We discuss the operational principle, present a short description of the manufacturing procedure, and report first measurements of effective gain and ion backflow.

## 2. M-THGEM with in-built meshes: manufacturing and operational principle

The MM-THGEM accommodates, in a single micro-structure, two very different gas avalanche concepts: the parallel-plate type micro-mesh and the hole-type (multi-layer THGEM) multiplier.

The Multi-layer Thick Gaseous Electron Multiplier (M-THGEM) is one of the newer members of the MPGD family [18], [19], consisting of a single, robust assembly of two/three THGEM elements laminated together. The large thickness of the insulator substrate results in a solid structure that provides large area coverage without the need of supporting frames, generally used with more fragile structures to avoid deformations caused by electrostatic forces. Upon the application of a voltage difference across the various outer and inner electrodes, a strong dipole field is established within the M-THGEM holes, which is responsible for a high electron collection efficiency and stable charge avalanche multiplication.

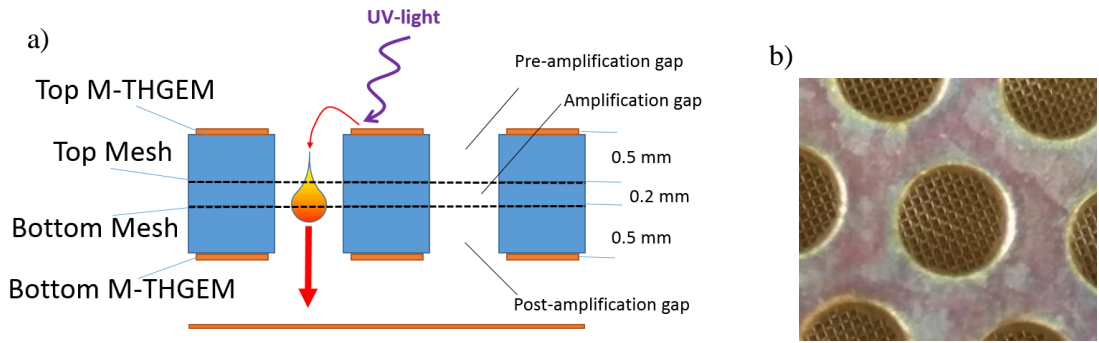

Figure 1. a) Schematic representation of the operational principle of the MM-THGEM multiplier. b) Color-enhanced photograph of the top surface and inner mesh electrodes of a MM-THGEM.

Different than the conventional multi-layer M-THGEM, the inner electrodes of the MM-THGEM consist of two micro-meter meshes to form a small (0.2 mm in this work) amplification gap (figure 1). The electrons are efficiently focused into the holes, transported into the small avalanche gap sandwiched between the two meshes, and therein multiplied by a strong uniform electric field (figure 2a). The electric field on the outer regions of the holes may also impart a pre- and post- electron amplification depending on the operational mode and the voltage bias configuration.

Due to the high field gradient between the amplification gap (delimited by the two meshes) and the upper pre-amplification region, a fraction of the ions produced in avalanche, as well as the ions coming from underneath multiplier elements, are trapped on the uppermost mesh (see figure 2b). Another large fraction of the avalanche ions are further trapped on the top surface of the MM-THGEM due to the dipole field geometry, while the rest migrates back to the drift volume. Figure 2b shows an example of electron/ion transport and avalanche simulation performed with Garfield/Magboltz [20], [21] package, which qualitatively illustrates the process of ion backflow reduction in MM-THGEM detector.

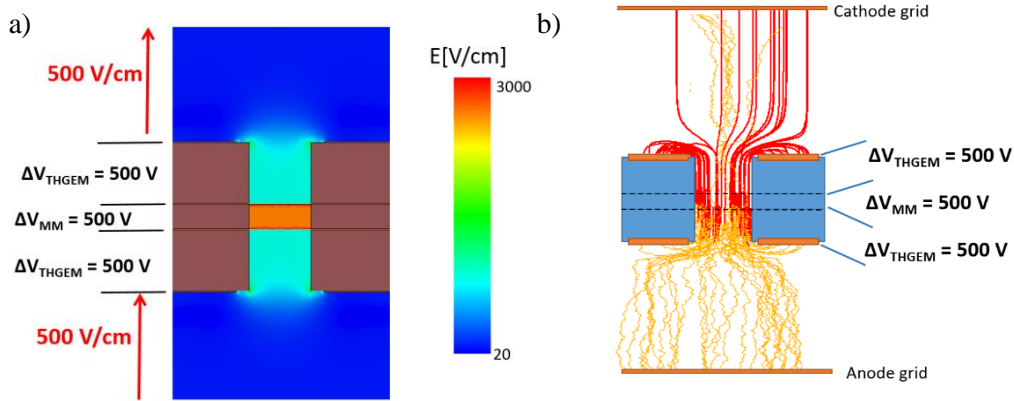

Figure 2. a) Map of the electric field strength in the MM-THGEM hole. b) Garfield/Magboltz simulation of the electron avalanche and ion drift lines. The computation in both a) and b) we carried out assuming a voltage difference of 500 Volt applied across the various (inner/outer) electrodes.

In the present work we have tested the performance of a first MM-THGEM prototype characterized by holes with a 0.7 mm diameter and a 1 mm spacing. The prototype is made out of three copper-cladded glass epoxy plates (EMC 370-5). The two outer insulator substrates are 0.5 mm thick while the inner substrate, which separates the two meshes and defines the avalanche gap, is 0.2 mm thick. The three boards (PCB1, PCB2 and PCB3 in figure 3) are mechanically drilled using the same approach of the conventional THGEM production. On the top surface of PCB1 and bottom surface of PCB3, small (0.1 mm) copper rims were etched around the holes by a photolithographic technique to reduce the discharge probability due to mechanical defects. Before the mechanical drilling, the inner surfaces of the PCB boards were laminated with a thin layer of B-stage epoxy glue (yellow layers in Figure 3), which allows for the fastening of all the electrodes in a solid structure. Two thin stainless steel meshes (Bopp 45/18) [22] were precisely cut to match the MM-THGEM size and then placed between the boards (Figure 3). Because of the holes' small diameter and the robust support of the insulator substrate, there is no need to stretch or frame the meshes, which were freely dropped over the PCB boards' surfaces. All the elements were then bound together (Figure 4) using a vacuum pressing system (up to 20 bar) operated at high temperature (180 $^{\circ}$C) to ensure a good uniformity across the effective area and close contact between all the various elements. Guiding pins (not included in the figures below) were used to prevent misalignment.

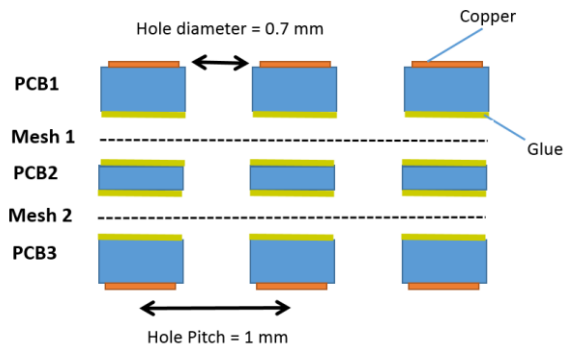

Figure 3. First stage of the MM-THGEM production consists on mechanical drilling and adhesive lamination of the PCB. The meshes are precisely cut to match the MM-THGEM effective area and positioned between the PCBs.

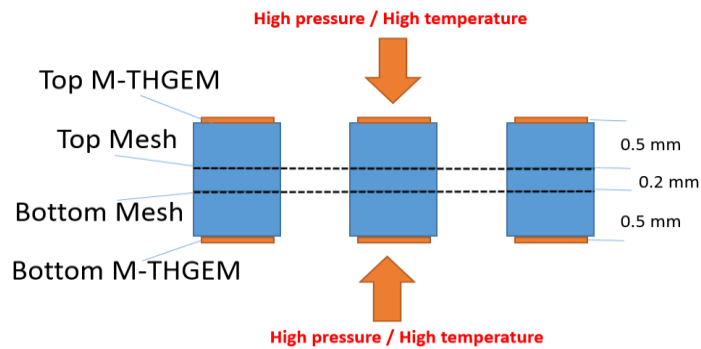

Figure 4. The MM-THGEM production is completed by bounding the PCB and the meshes using a vacuum press system operated at high temperature.

Other materials for the insulator substrate can be envisioned: for instance, Plexiglas could be a good alternative to the glass epoxy, since it does not required an intermediate B-stage epoxy glue layer. The number of the insulator substrates can be selected to create different geometries (up to a stack of 10 meshes) according to the application and operational requirements. The thickness of the insulator substrate can range from 0.1 mm up to a few mm (limited by the mechanical drilling stage). As in conventional Micromegas detectors, different mesh types can be used, including electroformed type meshes and the mesh pitch can range from 900 LPI down to 100 LPI.

## 3. Results

### 3.1 Effective gain

The MM-THGEM-based detector prototype (10×10 cm$^2$ effective area) was installed in a stainless-steel vessel, evacuated down to 10$^{-6}$ Torr before gas filling. During normal operation the vessel was flushed at a flow of 20-30 sccm. Data was taken after several hours of constant gas flow to avoid change in the gain due to outgassing of the detector components. The gain was measured in current mode [11] by illuminating the MM-THGEM top bare-Cu surface with a high-intensity UV-lamp through a quartz window, while the current collected on an anode board was measured with a high-resolution pico-ammeter. All the MM-THGEM electrodes were independently powered through a 5 MΩ resistor using modular high-voltage power supplies (Iseg model Mhn-30-250-vU CPS series [23]). The transfer field between the MM-THGEM electrode and the anode board (separated by 2 mm) was set to 500 V/cm (figure 1). The maximum achievable gain was defined by the onset of the anode current instabilities; the measurements and detector bias were discontinued after the appearance of the first discharges (generally occurring between the MM-THGEM meshes).

The gas gain was measured as the ratio of the current collected at the anode to the photoelectron current extracted from the MM-THGEM top electrode. The photocurrent (typically a few tens pA) was independently estimated by reversing the field on the drift region and collecting the charges on a cathode grid. The total avalanche current recorded at the anode was kept below the limit of 20 nA by means of filters placed between the UV-lamp and the quartz window.

The gain curves of the MM-THGEM was measured for different gas mixtures, He/(10%)$CO_2$ (Figure 5a) and in Ar/(10%)$CH_4$ (Figure 5b), and for different bias configurations. The latter is characterized by the parameter ($\alpha$), defined as the ratio of the potential difference applied across the amplification gap ($\Delta V_{MM}$ in figure 2a) to the potential difference applied across the pre- and post-amplification regions ($\Delta V_{THGEM}$ in figure 2a). As shown in figure 5, the higher maximum achievable gains (>$10^5$ and >$10^4$ for He/$CO_2$ and Ar/$CH_4$ respectively) were attained at large values of $\alpha$ ($\alpha$>0.75), which corresponds to the configurations where the avalanche processes are largely confined in the small volume between the two meshes with small or no significant contributions to the gain from the pre- and post-amplification regions. Stable higher gain maybe also possible for the lower overall operational voltages that are required in these conditions.

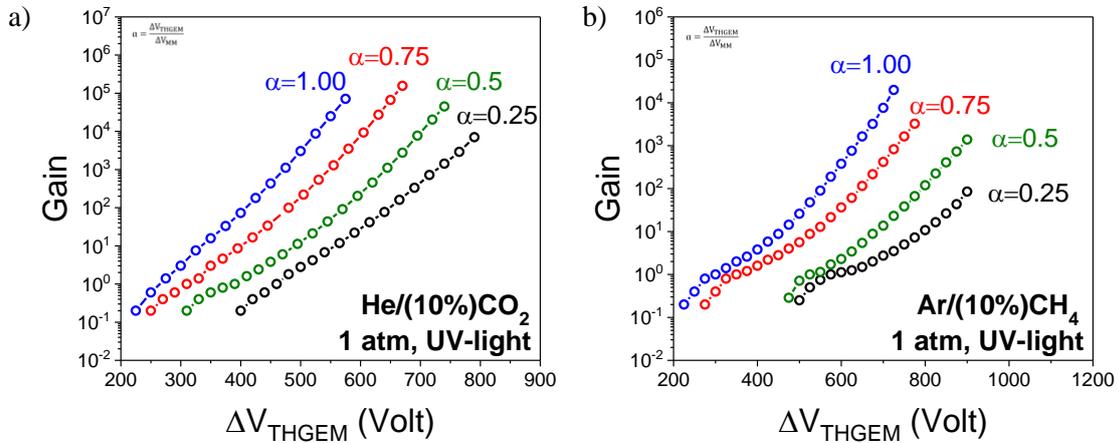

Figure 5. Gas gain versus voltage for single-electron detection in He/(10%)$CO_2$ and in Ar/(10%)$CH_4$, at atmospheric pressure.

## 3.2 Ion backflow (IBF)

The measurements of IBF in MM-THGEM-based detector were carried out with the experimental setup illustrated in figure 6. The MM-THGEM was coupled to a WELL-THGEM multiplier [24] in a two-cascade arrangement and operated in He/(10%)$CO_2$ at atmospheric pressure. The transfer gap between the WELL-THGEM and the MM-THGEM was 2 mm wide and the transfer field between the two electrodes was kept at a value of 500 V/cm. The two-stage detector arrangement allows for a larger detector gain, which is the product of the gas gain provided by each individual multiplier element. Maximum achievable gains in the range of $10^5$-$10^6$ were achieved depending on the detector bias configuration.

The ion and electron currents, collected at the cathode and at the anode respectively, were read out by two custom, fully floating, battery powered pico-ammeters characterized by a resolution at the 1 pA level [25]. The top surface of the MM-THGEM, illuminated by the UV-light through the quartz window, was electrically connected to an upper grid (5 mm apart) and biased thought the pico-ammeter.

In this work, the IBF is computed as the number of positive ions collected at the drift cathode ($I_{ions}$) normalized by the total number of electrons collected at the anode readout ($I_{elec}$). Notice that, in the case of single electron detection, the normalization to the current collected at the anode is equivalent to the normalization by the effective gain.

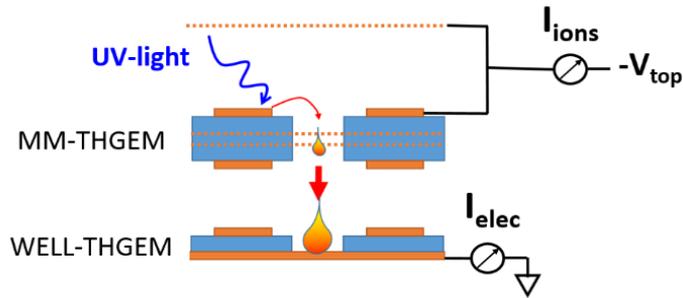

Figure 6. Scheme of the setup used to measure the IBF.

Figure 7b depicts the IBF as a function of the total detector gain, measured for three MM-THGEM operational modes, each of them characterized by different values of the parameter α (see definition in section 3.1). The total detector gains were progressively increased by varying the bias voltages of the MM-THGEM electrodes, while the voltage difference across the WELL-THGEM holes was kept fixed at 900 Volt (corresponding to a gas gain of around 20).

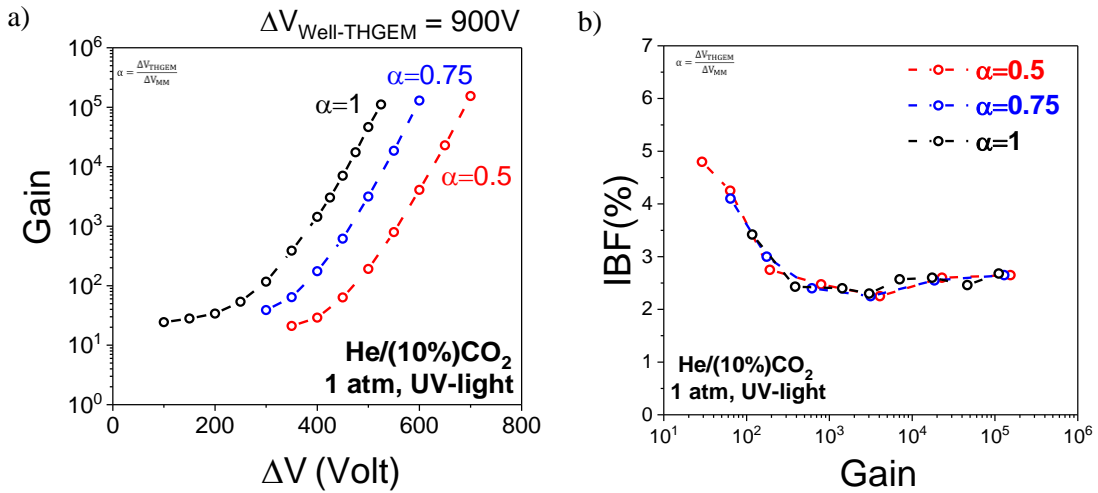

Figure 7. Gain curves (part a) and IBF as a function of the effective gain (part b) for different MM-THGEM bias configurations (expressed in term of the parameter α). In all the measurements, the voltage difference across the WELL-THGEM electrode was kept at 900 Volt, corresponding to a gas gain of around 20.

As shown in figure 7b, the IBF reaches a constant value of 2% when the gain is above 100 – this IBF is several time smaller compared to one attained with GEM/THGEM in an equivalent detector setup (two-cascade elements). Note that most of the gas gain is delivered by the MM-THGEM electrode, and therefore the IBF measurements carried out in these conditions are basically an indication of the capability of the MM-THGEM to trap ions generated by its own avalanches.

Similar methodology was used to investigate the capability of the MM-THGEM to trap ions generated by underneath multipliers (WELL-THGEM in this work) in a cascade detector setup. Figure 8b depicts the IBF values measured as a function of the total detector gain, where different biases were used to vary the gain sharing configurations between the two multipliers. For instance, the black curve in figure 8b shows the results of the IBF measurements when a fixed voltage difference of 1100 Volt was applied across the WELL-THGEM holes, which correspond to a WELL-THGEM contribution of 2000 to the total detector gain. The blue and red curves were obtained by applying a voltage difference of 1000 Volt and 900 Volt across the WELL-THGEM, which corresponds to a gain contribution of 400 and 20 respectively (Figure 8a).

The results depicted in figure 8b suggest that the MM-THGEM is extremely efficient in trapping ions coming for the lower transfer field; lower IBFs (ranging from 1.5-2%) are obtained when the contribution of the WELL-THGEM electrode to the total gain is high (black graph in figure 8b). This can be understood by considering that while the ions generated by avalanche within the MM-THGEM can be trapped only by the upper MM-THGEM mesh, ions streaming upwards from the underneath multipliers can also be trapped in the bottom mesh, though this process has a lower efficiency due to the negative electric field gradient.

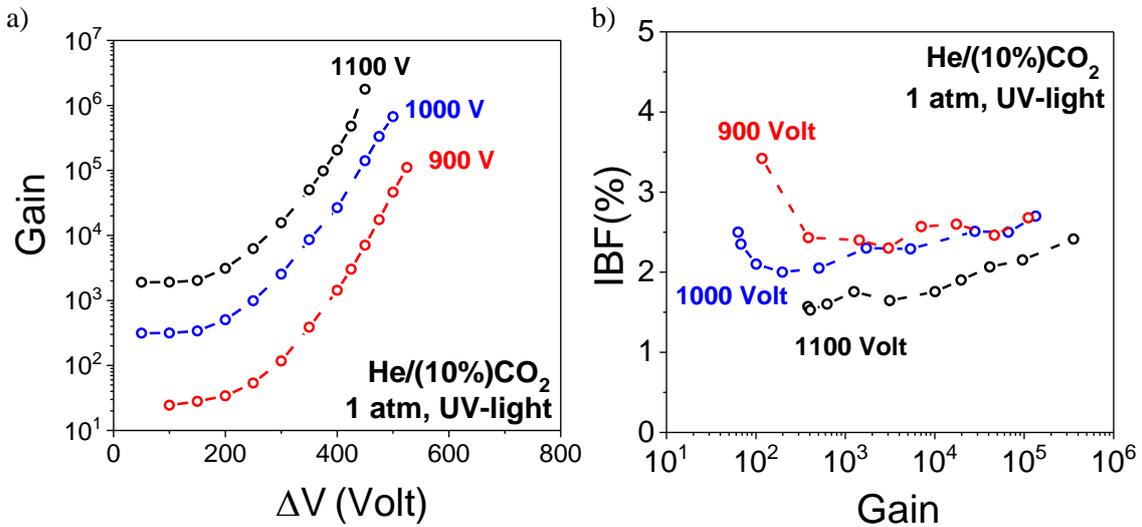

Figure 8. Gain curves (part a) and IBF as a function of the total detector gain (part b), computed for different detector bias configurations (i.e. different contribution of the WELL-THGEM to the total effective gain).

## 4. Conclusion

The operational principle and performance of a novel charge avalanche device, the multi-mesh THGEM (MM-THGEM), has been presented and discussed for the first time. The basic design of the MM-THGEM results from merging the two most popular, though conceptually dissimilar, MPGD architectures: the hole-type (GEM-like) multiplier and parallel-plate (Micromegas) avalanche structure. The MM-THGEM consists of two metallic micro-meshes separated by few hundred microns (0.2 mm in this work) inserted in a three-layer M-THGEM electrode. Ionization electrons from the drift region above the MM-THGEM are collected into the holes and transferred into the region between the two meshes where they are multiplied by a strong uniform electric field.

Notice that in a dipole field geometry, characteristic of hole-type multipliers (i.e. GEM/THGEM), the drifting electrons are subjected to an electric field strength that depends on their path inside the hole, which leads to a large range of possible avalanche multiplication factors. This results in a large fluctuation of the avalanche statistics, and thus a worse energy resolution in comparison to the parallel-plate geometry [26]. In this respect, the confinement of the avalanche in the uniform electric field established between the two MM-THGEM meshes might result in a good energy resolution compared to the conventional GEM/THGEM devices. On the other hand, the negative field gradient at the interface between amplification and post-amplification regions (see figure 2a) might cause a loss of electron transparency of the MM-THGEM bottom mesh, resulting in a degradation of the avalanche statistics. A systematic study of the energy resolution and of other operational properties of MM-THGEM detectors for different bias and geometries (including operation in a WELL configuration) is planned and the results will be presented elsewhere.

The design and parameter of the MM-THGEM, including the holes size and pitch, as well as the thickness of the insulator substrates can be tailored according to the applications requirements and operational conditions. This includes the possibility of assembling two or more elements in cascade which provides an increased stability, higher maximum achievable gains, and an overall lower IBF. Furthermore, the MM-THGEM can be assembled in a WELL (close-bottom) configuration: namely, the readout anode is placed in direct contact with the surface of the bottom MM-THGEM electrode, closing the MM-THGEM holes. This alternative mode of operation provides a robust double-mesh Micromegas geometry over large area. In the case of spark-induced permanent damages of the meshes or of the insulator substrates, the MM-THGEM can easily be dismounted from the readout board and swapped with a new electrode. Note that, in the case of single-/double-mesh Micromegas directly populated over the readout board, the recovering is extremely unlikely and thus the full assembly has to be replaced. Furthermore, by operating the MM-THGEM in a WELL configuration, the negative field gradient at the interface between the amplification and post-amplification regions can be cancelled out, leading in principle to a better energy resolutions.

We have performed a first characterization of the operation properties of the MM-THGEM in standard gas mixtures at atmospheric pressure. Gains above $10^5$, $10^4$ for single-electron detection, were achieved with a single element in He/(10%)$CO_2$ and Ar/(10%)$CH_4$ respectivelly. In a double-cascade MM-THGEM/WELL-THGEM detector assembly the maximum achievable gain for single-electron detection reaches values above $10^6$ in He/(10%)$CO_2$.

One of the immediate advantages of the presence of the two meshes in the MM-THGEM structure is a high capability to capture the positive ions. In a double-cascade MM-THGEM/WELL-THGEM detector we measured IBFs in the range of 1.5-2%, at a gain of $10^3$-$10^6$. These values are several times lower compared to a similar double GEM/THGEM detectors. Moreover, the above IBF values were attained for single photoelectron detection where the top MM-THGEM surface and the upper cathode grip were connected together (zero drift field). On one hand, this "reflective" gaseous photomultiplier configuration provides the higher photo-electron collection efficiency [11]. On the other hand, all the ions that escape the meshes are collected either at the MM-THGEM top surface or on the upper cathode grid and all of them contribute to the IBF. On the contrary, in a typical TPC configuration characterized by a non-zero drift field, the $I_{ions}$ is defined only by the fraction of ions that migrate all the way up to the cathode grid, while a large fraction of ions is further collected on the MM-THGEM top surface. As a result, lower IBFs (< 1%) may be expected in TPC mode with a cascade of several MM-THGEM electrodes.

**References**

[1] S. Dalla Torre, "Status and perspectives of gaseous photon detectors," *Nucl. Instrum. Methods Phys. Res. Sect. Accel. Spectrometers Detect. Assoc. Equip.*, vol. 639, no. 1, pp. 111–116, May 2011.
[2] R. Chechik and A. Breskin, "Advances in gaseous photomultipliers," *Nucl. Instrum. Methods Phys. Res. Sect. Accel. Spectrometers Detect. Assoc. Equip.*, vol. 595, no. 1, pp. 116–127, Sep. 2008.
[3] D. Attié, "TPC review," *Nucl. Instrum. Methods Phys. Res. Sect. Accel. Spectrometers Detect. Assoc. Equip.*, vol. 598, no. 1, pp. 89–93, Jan. 2009.
[4] D. Suzuki et al., "Prototype AT-TPC: Toward a new generation active target time projection chamber for radioactive beam experiments," *Nucl. Instrum. Methods Phys. Res. Sect. Accel. Spectrometers Detect. Assoc. Equip.*, vol. 691, pp. 39–54, Nov. 2012.
[5] A. Breskin, "CsI UV photocathodes: history and mystery," *Nucl. Instrum. Methods Phys. Res. Sect. Accel. Spectrometers Detect. Assoc. Equip.*, vol. 371, no. 1–2, pp. 116–136, Mar. 1996.
[6] F. Sauli, L. Ropelewski, and P. Everaerts, "Ion feedback suppression in time projection chambers," *Nucl. Instrum. Methods Phys. Res. Sect. Accel. Spectrometers Detect. Assoc. Equip.*, vol. 560, no. 2, pp. 269–277, May 2006.
[7] E. Nappi and V. Peskov, *Imaging Gaseous Detectors and Their Applications*. .
[8] F. Sauli, "GEM: A new concept for electron amplification in gas detectors," *Nucl. Instrum. Methods Phys. Res. Sect. Accel. Spectrometers Detect. Assoc. Equip.*, vol. 386, no. 2–3, pp. 531–534, Feb. 1997.
[9] Y. Giomataris, P. Rebourgeard, J. P. Robert, and G. Charpak, "MICROMEGAS: a high-granularity position-sensitive gaseous detector for high particle-flux environments," *Nucl. Instrum. Methods Phys. Res. Sect. Accel. Spectrometers Detect. Assoc. Equip.*, vol. 376, no. 1, pp. 29–35, Jun. 1996.
[10] P. Bhattacharya et al., "Investigation of ion backflow in bulk micromegas detectors," *J. Instrum.*, vol. 10, no. 09, pp. P09017–P09017, Sep. 2015.
[11] C. Shalem, R. Chechik, A. Breskin, and K. Michaeli, "Advances in Thick GEM-like gaseous electron multipliers—Part I: atmospheric pressure operation," *Nucl. Instrum. Methods Phys. Res. Sect. Accel. Spectrometers Detect. Assoc. Equip.*, vol. 558, no. 2, pp. 475–489, Mar. 2006.
[12] F. Sauli, "The gas electron multiplier (GEM): Operating principles and applications," *Nucl. Instrum. Methods Phys. Res. Sect. Accel. Spectrometers Detect. Assoc. Equip.*, vol. 805, pp. 2–24, Jan. 2016.
[13] M. Alexeev et al., "Ion back flow reduction in a THGEM based detector," in *2012 IEEE Nuclear Science Symposium and Medical Imaging Conference (NSS/MIC)*, 2012, pp. 1165–1171.
[14] J. F. C. A. Veloso, J. M. F. dos Santos, and C. A. N. Conde, "A proposed new microstructure for gas radiation detectors: The microhole and strip plate," *Rev. Sci. Instrum.*, vol. 71, no. 6, pp. 2371–2376, Jun. 2000.
[15] A. Lyashenko, A. Breskin, R. Chechik, J. M. F. dos Santos, F. D. Amaro, and J. F. C. A. Veloso, "Efficient ion blocking in gaseous detectors and its application to gas-avalanche photomultipliers sensitive in the visible-light range," *Nucl. Instrum. Methods Phys. Res. Sect. Accel. Spectrometers Detect. Assoc. Equip.*, vol. 598, no. 1, pp. 116–120, Jan. 2009.
[16] F. D. Amaro, C. Santos, J. F. C. A. Veloso, A. Breskin, R. Chechik, and J. M. F. dos Santos, "The Thick-COBRA: a new gaseous electron multiplier for radiation detectors," *J. Instrum.*, vol. 5, no. 10, pp. P10002–P10002, Oct. 2010.
[17] F. Jeanneau, M. Kebbiri, and V. Lepeltier, "Ion back-flow gating in a micromegas device," *Nucl. Instrum. Methods Phys. Res. Sect. Accel. Spectrometers Detect. Assoc. Equip.*, vol. 623, no. 1, pp. 94–96, Nov. 2010.

[18] M. Cortesi *et al.*, "Multi-layer thick gas electron multiplier (M-THGEM): A new MPGD structure for high-gain operation at low-pressure," *Rev. Sci. Instrum.*, vol. 88, no. 1, p. 013303, Jan. 2017.

[19] Y. Ayyad, M. Cortesi, W. Mittig, and D. Bazin, "$CO_2$ operation of an active target detector readout based on THGEM.," *J. Instrum.*, vol. 12, no. 06, pp. P06003–P06003, Jun. 2017.

[20] F. V. Böhmer *et al.*, "Simulation of space-charge effects in an ungated GEM-based TPC," *Nucl. Instrum. Methods Phys. Res. Sect. Accel. Spectrometers Detect. Assoc. Equip.*, vol. 719, pp. 101–108, Aug. 2013.

[21] R. Veenhof, *Garfield - simulation of gaseous detectors*. 1984-2010 (http://garfield.web.cern.ch).

[22] Saati company, "BOPP™ Stainless Stell Wire Cloth (datasheet) (http://www.saati.com/)." .

[23] ISEG GmbH, "Modular High Voltage Power Supply CPS Series (https://iseg-hv.com/en/products/detail/CPS)." .

[24] L. Arazi, M. Pitt, S. Bressler, L. Moleri, A. Rubin, and A. Breskin, "Laboratory studies of THGEM-based WELL structures with resistive anode," *J. Instrum.*, vol. 9, no. 04, p. P04011, Apr. 2014.

[25] S. Dalla Torre, B. Gobbo, S. Levorato, G. Menon, and F. Tessarotto, "RHIP, a radio-controlled high-voltage insulated picoammeter," in *Nuclear Science Symposium, Medical Imaging Conference and Room-Temperature Semiconductor Detector Workshop (NSS/MIC/RTSD), 2016*, 2016, pp. 1–3.

[26] T. Zerguerras, B. Genolini, J. Peyré, J. Pouthas, and V. Lepeltier, "MPGD's spatial and energy resolution studies with an adjustable point-like electron source," *Nucl. Instrum. Methods Phys. Res. Sect. Accel. Spectrometers Detect. Assoc. Equip.*, vol. 581, no. 1–2, pp. 258–260, Oct. 2007.